# Throwing and jumping for maximum horizontal range

Nicholas P. Linthorne[a)]
*School of Sport and Education, Brunel University, Uxbridge, Middlesex, UB8 3PH, United Kingdom*

Optimum projection angles for achieving maximum horizontal range in throwing and jumping events are considerably less than 45°. This unexpected result arises because an athlete can generate a greater projection velocity at low projection angles than at high angles. The range of a projectile is strongly dependent on projection speed and so the optimum projection angle is biased towards low projection angles. Here we examine the velocity-angle relation and the optimum projection angle in selected throwing and jumping events.

## I. INTRODUCTION

Several throwing and jumping activities involve projecting a ball, implement, or the human body for maximum horizontal range. One might expect that the optimum projection angle for these activities would be about 45°. However, video measurements have revealed that sports projectiles are usually launched at angles much lower than 45° (Table I).

Table I. Typical projection angles in throwing and jumping events.

| Event | Angle (°) | Reference |
|---|---|---|
| shot put | 26 – 41 | 1 |
| javelin throw | 26 – 40 | 2 |
| discus throw | 27 – 43 | 3 |
| hammer throw | 37 – 44 | 4 |
| long jump | 15 – 27 | 5 |
| standing long jump | 25 – 31 | 6 |
| soccer throw-in | 23 – 37 | 7 |

Aerodynamic forces and the height difference between the launch and landing are not the main causes of the low projection angles. For all the events listed in Table I the height difference between the launch and landing reduces the optimum projection angle by no more than a few degrees.[8,9] Although aerodynamic lift and drag can substantially affect the range of a sports projectile, the effect on the optimum projection angle is relatively small. Aerodynamics forces reduce the optimum projection angle in the javelin throw, discus throw, and soccer throw-in by only a few degrees.[7,10,11] In the hammer throw the effect is less than one degree and there is almost no effect in the shot put, long jump, and standing long jump.[6,9,12,13]

The main reason for the low projection angles in throwing and jumping is that the projection velocity an athlete can produce decreases with increasing projection angle.[14] The musculoskeletal structure of the human body is such that an athlete can generate a greater projection velocity in the horizontal direction than in the vertical direction. Because the range of a projectile is strongly dependent on the projection velocity, even a small dependence of projection velocity on projection angle is sufficient to lower the optimum projection angle to substantially below 45°.

The present article summarizes recent work on the optimum projection angles in the shot put, long jump, standing long jump, and soccer throw-in.[15–18] We used video measurements to experimentally determine the dependence of projection velocity on the angle of projection.[19] The athlete's optimum projection angle was calculated by substituting the velocity-angle relation into the equations for the flight trajectory of the projectile. This method produced good agreement between the calculated optimum projection angles and the athletes' preferred projection angles. The principles presented in this article have application to all throwing and jumping events in which the aim is to project a sports projectile for maximum horizontal range. For further details of the measurement techniques, mathematical models, experimental results and uncertainties, the interested reader should consult the original papers.[14–18]

Projectile motion is one of the staple topics in a mechanics course, and educators often use sporting examples in an effort to capture the student's interest. The shot put and long jump are suitable for discussion in an introductory undergraduate class as the sports projectile can be assumed to be in free flight. The soccer throw-in, hammer throw, javelin throw, and discus throw require an aerodynamic treatment and are more suited for advanced classes. In our experience, many students find research projects in which video measurements are performed on athletes to be particularly engaging.

## II. SHOT PUT

The shot put (Fig. 1) is the best-known example in which a sports implement behaves as a projectile in free flight. When a projectile is launched from ground level over a horizontal plane the range of the projectile is given by $R = (v^2 \sin 2\theta)/g$, where $v$ is the projection velocity, $\theta$ is the projection angle, and $g$ is the acceleration due to gravity. In shot-putting the athlete launches the shot from above ground level and so the range is given by;

$$R = \frac{v^2 \sin 2\theta}{2g}\left[1 + \left(1 + \frac{2gh}{v^2 \sin^2\theta}\right)^{1/2}\right], \quad (1)$$

where $h$ is the height difference between launch and landing. The optimum projection angle may be determined by graphical techniques or by differentiation. For example, if an athlete projects the shot from 2.1 m above the ground at a projection velocity of 13 m/s, the optimum projection angle is 42°.[9] However, this calculated optimum projection angle is considerably greater than the projection angles used by actual athletes (26–41°).

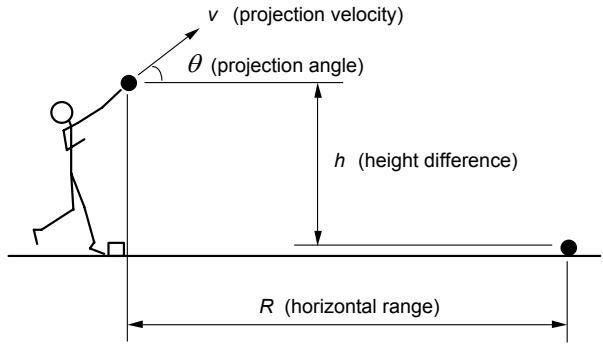

Fig. 1. Diagram of shot-putting showing the release parameters that determine the horizontal range of the shot. Adapted with permission.

### A. Projection velocity

The error in the above method of determining the optimum projection angle is that we assumed the projection velocity $v$ is independent of the projection angle $\theta$. Our experiments on shot-putters have shown that an athlete can project the shot faster at low projection angles than at high projection angles. Figure 2 shows data for a male college shot-putter.[15] To determine the athlete's optimum projection angle we must obtain a mathematical expression for the athlete's relation between projection velocity and projection angle, $v(\theta)$, and then substitute this expression into Eq. (1). The result is shown in Fig. 3. Note that the calculated optimum projection angle is now much less than 42°.

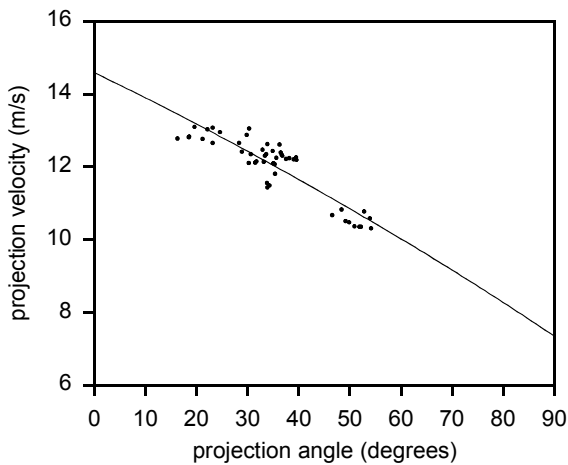

Fig. 2. Decrease in projection velocity with increasing projection angle for a male shot-putter. Adapted with permission.

In an introductory class on projection angles in sports the student can generate Fig. 3 using a spreadsheet and graphing program such as *Microsoft Excel*. Asking the student to generate the dashed curves in Fig. 3 is a useful exercise to illustrate how the conventional (but incorrect) optimum projection angle of just under 45° is obtained. The student can be given the mathematical expression for $v(\theta)$, or it can be obtained by fitting an appropriate curve to some real data (obtained in a student research project). An appropriate fit to some velocity-angle data may be as simple as a linear expression, but a better approach is to use an expression that is based on a mathematical model of shot-putting, such as that described below.

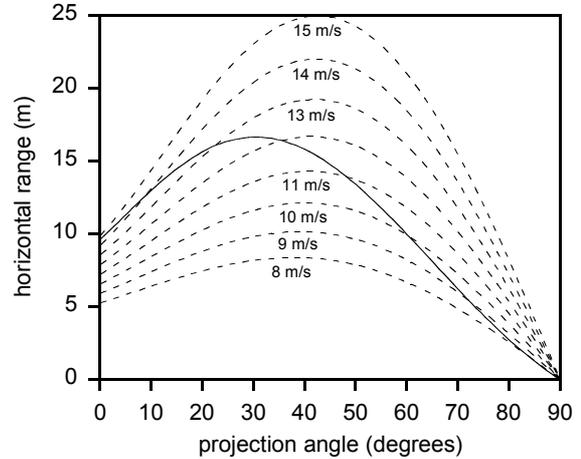

Fig. 3. Range of a shot (solid line) calculated by substituting the observed relation between projection velocity and projection angle for the athlete (Fig. 2) into the equation for the range of a projectile in free flight [Eq. (1)]. The optimum projection angle for this athlete is about 31°. The dashed lines are the calculated range of a shot, assuming constant values of projection velocity and a height difference of $h$ = 2.1 m. Adapted with permission.

### B. Model of projection velocity

The projection phase of the shot put may be modeled by assuming the athlete applies a constant force $F$ to the shot.[15] This force accelerates the shot from rest along a straight line path $l$ to produce a projection velocity $v$. Applying the law of conservation of energy to the projection phase gives an expression for the projection velocity;

$$v = \sqrt{\frac{2Fl}{m}}, \qquad (2)$$

where $m$ is the mass of the shot (7.26 kg for men, and 4.00 kg for women).

Unfortunately, Eq. (2) does not agree with the experimental data (Fig. 2) as it says that the projection velocity is the same at all projection angles. In the above model we assumed that the force exerted by the athlete on the shot is the same for all projection angles, but in practice the human body can produce a greater throwing force in the horizontal direction than in the vertical direction. Assuming the force exerted by the athlete on the shot decreases linearly with projection angle, we obtain;

$$v(\theta) = \sqrt{\frac{2(F_o - a\theta)l}{m}}, \qquad (3)$$

where $F_o$ is the average force exerted on the shot for a horizontal projection angle, and $a$ is a constant that characterizes the athlete's force decrease with increasing projection angle.

Equation (3) gives a good fit to the experimental data. For the male shot-putter shown in Fig. 2 the acceleration path length is $l = 1.65$ m, the average force exerted for a horizontal throw is $F_o = 460$ N, and the rate of decrease in force is $a = 4.1$ N/degree. The calculated values of $l$ and $F_o$ agree with values obtained from a video analysis of world-class male shot-putters, and the value of $a$ is similar to that expected from the difference in weight lifted between a bench press exercise (160 kg at $\theta \approx 0°$) and a shoulder press exercise (120 kg at $\theta \approx 90°$). A more accurate model of the projection phase would include the effect of gravity on the motion of the shot. However, in shot-putting $F_o \gg mg$ and so the weight of the shot has only a small influence on the athlete's projection angle.

A student can use the model presented here to investigate the effects of the athlete's strength and throwing technique on the maximum range and the optimum projection angle. The average force $F_o$ is related to the athlete's overall strength, and $a$ is some complex function of the athlete's throwing technique and the relative strengths of the muscles used in the throwing movement. Male athletes have values of $F_o$ of between 100 and 800 N, and $a$ lies between 2 and 5 N/degree. An athlete's maximum range is mostly determined by the value of $F_o$, whereas the optimum projection angle is mostly determined by the value of $a$. The greater the rate of decrease in velocity, the lower the athlete's optimum projection angle.

**C. Projection height**

An astute student will notice that in Eq. (1) the height difference $h$ may also be a function of the projection angle. Figure 4 shows projection height data for a male college shot-putter.[15] The increase in projection height with increasing projection angle arises from the configuration of the athlete's body at the instant of release (Fig. 5). When using a high projection angle the angle of the athlete's arm to the horizontal is greater, and therefore at the instant of release the shot is at a greater height above the ground. The relation between the height difference and the projection angle is given by

$$h(\theta) = h_{shoulder} + l_{arm}\sin\theta - r_{shot}, \quad (4)$$

where $h_{shoulder}$ is the height of the athlete's shoulders when standing upright, $l_{arm}$ is the length of the athlete's outstretched throwing arm and shoulder, and $r_{shot}$ is the radius of the shot (6 cm for men, and 5 cm for women). The values of $h_{shoulder}$ and $l_{arm}$ obtained from the fitted curve in Fig. 4 (1.68 m and 0.87 m) are in reasonable agreement with actual body dimensions.

The effect of the projection height on the optimum projection angle is calculated by substituting the expression for $h(\theta)$ into Eq. (1), along with the expression for the projection velocity $v(\theta)$ obtained previously. In an introductory class on projection angles in sports the student can be given the mathematical expression for $h(\theta)$, or it can be obtained by fitting Eq. (4) to some real data. For most people, the height of the shoulders when standing upright ($h_{shoulder}$) and the length of the outstretched arm and shoulder ($l_{arm}$) are about 82% and 52% of their standing height. The student will find that unlike the relation for projection velocity $v(\theta)$, the relation for projection height $h(\theta)$ has little influence on the athlete's optimum projection angle.

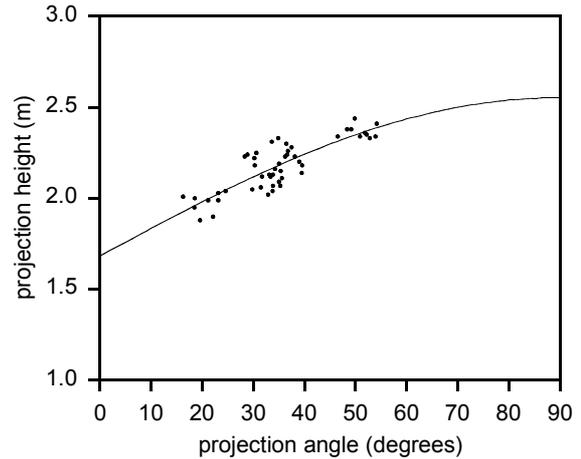

Fig. 4. Increase in projection height for a male shot-putter. Adapted with permission.

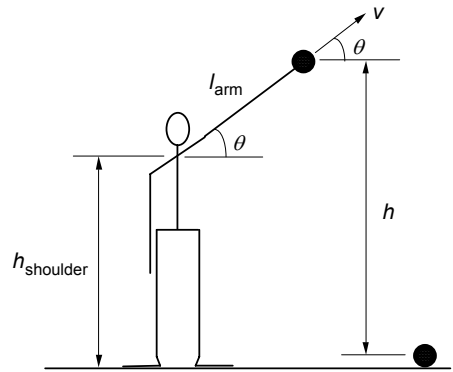

Fig. 5. Anthropometric model of a shot-putter at the instant of releasing the shot. Adapted with permission.

**D. Video measurements of shot-putters**

An experimental study of the optimum projection angle in the shot put makes an engaging student research project.[19] A domestic video camera operating at 25 or 30 Hz is suitable for recording sports movements, but you may have to invest in some decent biomechanical analysis software. We use *APAS* software[20] as it is able to separate the two fields that make up each video frame and hence double the video sampling rate to 50 or 60 Hz. *APAS* has a module for smoothing the data using a digital filter, and the most appropriate cut-off frequency for the filter is selected by examining the power spectrum of the video data. The student will also learn how to calibrate a video image so as to convert pixel coordinates on a video image into real-world coordinates. In a student research project the calculated optimum projection angle may be compared to the athlete's preferred projection angle or to the measured throw distances (Fig. 6).

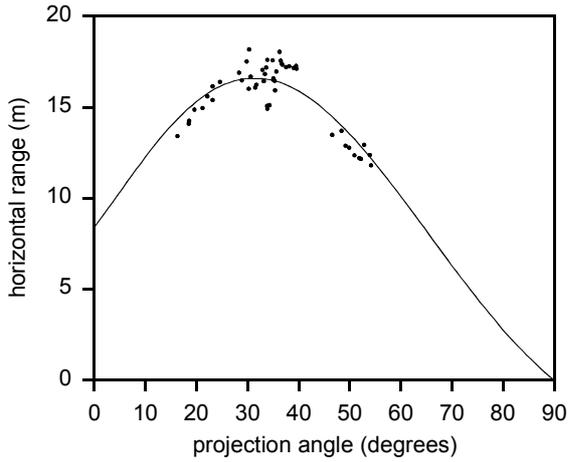

Fig. 6. Experimental data and calculated horizontal range for a male shot-putter.

### III. LONG JUMP

In jumping, the projectile is the human body, rather than a sports implement or sports ball. A long jumper in flight has a low drag-to-weight ratio and so acts like a projectile in free flight. We can therefore use the same technique for calculating the optimum projection angle as that used for the shot put.

Figure 7 shows the velocity-angle relation for a male long jumper.[16] In long jumping, the projection velocity is produced through a combination of horizontal velocity developed in the run-up and vertical velocity generated during the take-off. The highest projection velocities are obtained when the jumper uses a fast run-up and then attempts to jump up as much as possible. However, long jumpers cannot attain projection angles greater than about 25° using this technique. To achieve greater projection angles the athlete must use a slower run-up and so the projection velocity is reduced. In the extreme situation of a near-vertical projection angle, the run-up velocity must be reduced to walking pace and so the projection velocity is at its lowest.

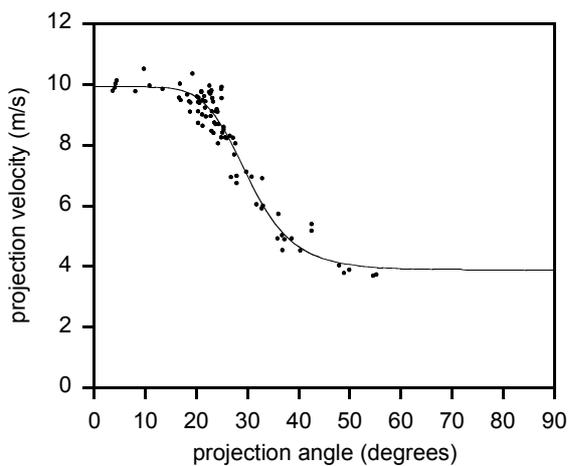

Fig. 7. Decrease in projection velocity for a male long jumper. Adapted with permission.

We have not been able to develop a simple mathematical model of jumping that reproduces the observed relation between projection velocity and projection angle (Fig. 7), but a symmetric logistic function produces a good empirical fit to the data;[16]

$$v(\theta) = \frac{v_{max} - v_{min}}{1 + \left(\frac{\theta}{\theta_{inf}}\right)^a} + v_{min}, \qquad (5)$$

where $v_{max}$ is the asymptotic maximum projection velocity (as when running straight through the take-off, where $\theta \approx 0°$), $v_{min}$ is the asymptotic minimum projection velocity (as in an upwards vertical jump from one leg, where $\theta \approx 90°$), $\theta_{inf}$ is the value of the projection angle at the inflection point of the curve, and $a$ is the slope coefficient. For the male long jumper shown in Fig. 7, $v_{max}$ = 9.9 m/s, $v_{min}$ = 3.9 m/s, $\theta_{inf}$ = 30°, and $a$ = 6.8.

Our measurements show that although the take-off and landing heights both increase with increasing projection angle, the height difference between the two remains approximately constant at about $h$ = 0.5 m. The athlete's optimum projection angle is calculated by substituting the expressions for $v(\theta)$ and $h(\theta)$ into Eq. (1). Figure 8 shows that the calculated optimum projection angle is about 21°, which is in good agreement with the angle used by this athlete in competition.

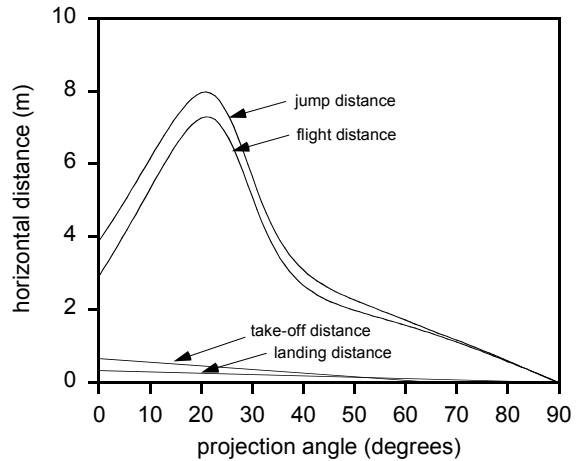

Fig. 8. Calculated jump distance for a male long jumper. Adapted with permission.

In long jumping the total jump distance is slightly more than the horizontal distance traveled through the air (the "flight distance"). The athlete's center of mass at the instant of take-off is ahead of the take-off line, and the center of mass at the instant of landing is behind the mark made in the sand by the jumper's feet (Fig. 9). Both the take-off and landing distances depend on configuration of the athlete's body, which in turn depend on the projection angle. The observed dependence of the take-off and landing distances on the projection angle are shown in Fig. 8. However, these component distances make relatively small contributions to the total jump distance and have little effect on the optimum projection angle.

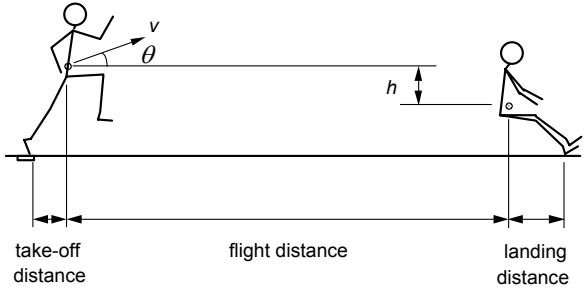

Fig. 9. In a long jump the jump distance is the sum of the take-off, flight, and landing distances. Adapted with permission.

A student research project on the optimum projection angle in the long jump is more complex than for the shot put. The *APAS* software can calculate the location of the center of mass of the human body (using representative data for the masses of the body segments). However, this requires the student to digitize 18 body points on each video image and is a time consuming process. A less tedious alternative is to use the athlete's hips as an estimate of the location of the athlete's center of mass.

## IV. SOCCER THROW-IN

A soccer ball is a moderately aerodynamic projectile, and when kicked or thrown the range of the ball is substantially reduced by aerodynamic drag. The optimum projection angle in a throw or kick may be calculated as before, but we must use aerodynamic equations to calculate the flight trajectory of the ball. We conducted a study of the optimum projection angle in the soccer throw-in where the ball was deliberately projected with little or no spin so as to eliminate the confounding effects of aerodynamic lift. Figure 10 shows the velocity-angle relation for a male athlete.[18]

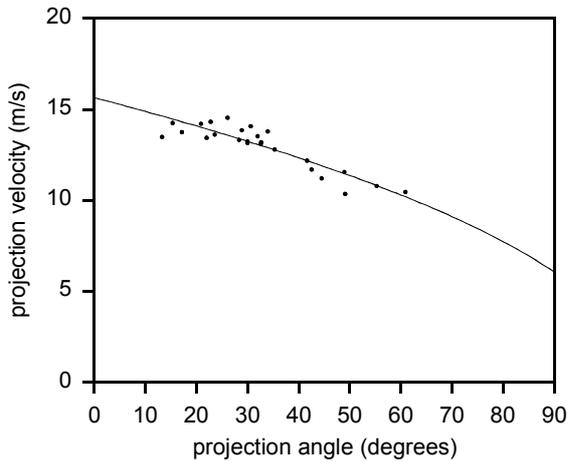

Fig. 10. Decrease in projection velocity for a male athlete throwing a soccer ball. Adapted with permission.

The projection phase of the throw may be modeled by assuming the athlete accelerates the ball by exerting a constant force $F$ over a path length $l$.[18] The weight of the ball ($m = 0.43$ kg) is negligible in comparison to the throwing force exerted by the athlete on the ball, and therefore the relation between the projection velocity and projection angle is mostly determined by the musculoskeletal structure of the human body. We assume that the force exerted by the athlete on the ball decreases linearly with projection angle. The expression for the relation between projection velocity and projection angle is then given by Eq. (3). For the fitted curve shown in Fig. 10 the acceleration path length is $l = 1.14$ m, the average force exerted for a horizontal throw is $F_o = 46$ N, and the rate of decrease in force is $a = 0.44$ N/degree.

In a soccer throw-in the projection height of the ball increases slightly with increasing projection angle, from about 2.0 m in a horizontal throw to about 2.4 m in a vertical throw. As in the shot put, this relation arises from the geometry of the throwing action.

The flight trajectory equations of the soccer ball are;[18]

$$\frac{d^2x}{dt^2} = -kv\left(C_D\frac{dx}{dt} + C_L\frac{dy}{dt}\right) \qquad (6)$$

$$\frac{d^2y}{dt^2} = kv\left(C_L\frac{dx}{dt} - C_D\frac{dy}{dt}\right) - g, \qquad (7)$$

where $v$ is the instantaneous velocity of the ball relative to the air. The constant $k$ is given by $k = \rho S/(2m)$, where $\rho$ is the air density (1.225 kg/m$^3$ at sea level and 15°C), and $S$ is the cross-sectional area of the ball (0.038 m$^2$). At speeds typical of the soccer throw-in, a soccer ball has a drag coefficient of about $C_D = 0.2$, and for a ball that is projected with zero spin the lift coefficient is about $C_L = 0$.

The initial conditions for the flight trajectory equations are generated from the athlete's expressions for $v(\theta)$ and $h(\theta)$. The flight trajectory equations are nonlinear, and so we used the numerical solution capabilities of *Mathematica* to calculate the flight trajectories. For the athlete in this study the optimum projection angle was about 30° (Fig. 11). As in the other throwing and jumping events discussed here, the projection velocity relation $v(\theta)$ has a strong effect on the optimum projection angle whereas the projection height relation $h(\theta)$ has almost no effect.

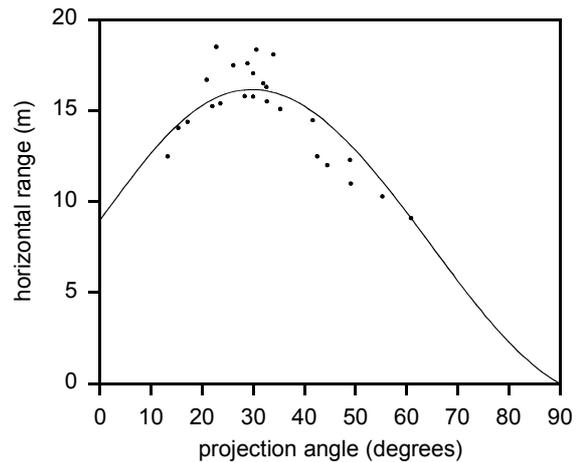

Fig. 11. Experimental data and calculated horizontal range for a male athlete throwing a soccer ball. Adapted with permission.

## V. OTHER EVENTS

### A. Standing long jump

The standing long jump is no longer contested in athletics meets, but it is still used as a test of leg power, particularly for athletes in sports that involve running and jumping. In the standing long jump the athlete's projection velocity decreases with increasing projection angle, mostly because the athlete must overcome an increasing fraction of his body weight.[17] The athlete's projection height and landing height both increase with increasing projection angle, but the height difference between the two remains approximately constant at about $h = 0.2$ m. Substituting the athlete's expressions for $v(\theta)$ and $h(\theta)$ into the equation for a projectile in free flight [Eq. (1)] gives a projection angle that maximizes the flight distance of about 35°. In a standing long jump the take-off and landing distances make relatively large contributions to the total jump distance and the optimum projection angle for the total jump is about 25°.

### B. Hammer throw

We have not yet determined the optimum projection angle in the hammer throw using the techniques presented in this article. The mass of a hammer is the same as a shot (7.26 kg for men, and 4.00 kg for women), but aerodynamic drag is not quite negligible in the hammer throw. Typical projection velocities in the hammer throw ($v \approx 25$ m/s) are about twice those in the shot put, and the drag area of the hammer is about twice as great as the shot because of the attached handle and wire.[21] Aerodynamic drag in the hammer throw reduces the range by about 5% and reduces the optimum projection angle by just under 1º. To the best of our knowledge there is no published information on the velocity-angle relation for throwing a hammer. We suspect a relatively weak dependence because measured projection angles are only a little below 45º.

### C. Javelin throw and discus throw

An analysis of the optimum projection angle in the javelin throw or discus throw is probably only suitable for an advanced student project as the aerodynamic flight equations are relatively complex.[11,14] The flight trajectory of a javelin depends on the angle of attack, angle of yaw, rate of pitch, and rate of spin, and these parameters change during the flight. Likewise, the flight of a discus is affected by the angle of attack and rate of spin. The velocity-angle relation for throwing a javelin is to be found in Ref. 14, but there is no corresponding information for throwing a discus.

[a]Electronic mail: nick.linthorne@brunel.ac.uk